%% file: main.tex
\documentclass[runningheads]{llncs}
\usepackage{graphicx}
\usepackage{amsmath}
\usepackage{listings}
\usepackage{xcolor}
\usepackage{lstautogobble}
\usepackage{hyperref}
\usepackage{cleveref}
\usepackage{wrapfig}
\usepackage{subcaption}
\usepackage{stmaryrd}
\usepackage{wrapfig}
\usepackage{mathpartir}
\usepackage{mathtools}
\usepackage{makecell}
\usepackage[inline]{enumitem}
\usepackage{xspace}
\usepackage{orcidlink}
\usepackage[misc]{ifsym}
\usepackage{booktabs}
\usepackage{algpseudocode}
\usepackage{algorithm}
\usepackage{wrapfig}
\usepackage{float}

\usepackage{amssymb}
\usepackage{todonotes}

\input{commands}

\begin{document}
\title{Cutting Corners on Uncertainty: Zonotope Abstractions for Stream-based Runtime Monitoring\thanks{This work was partially supported by the German Research Foundation (DFG) as part of PreCePT (FI 936/7-1; FR 2715/6-1) and by the European Research Council (ERC) Grant HYPER (No. 101055412).}}
\titlerunning{Cutting Corners on Uncertainty}
%
\author{Bernd Finkbeiner\inst{1,2}\orcidlink{0000-0002-4280-8441} \and
Martin Fränzle\inst{3}\orcidlink{0000-0002-9138-8340} \and
Florian Kohn$^\text{(\Letter)}$\inst{1}\orcidlink{0000-0001-9672-2398} \and\\
Paul Kröger\inst{3}\orcidlink{0000-0002-0301-3611}}
\authorrunning{Finkbeiner et al.}
%
\institute{CISPA Helmholtz Center for Information Security, Saarbrücken, Germany \email{\{finkbeiner,florian.kohn\}@cispa.de} \and
Technical University Munich, Germany \and
Carl von Ossietzky Universität, Oldenburg, Germany
\email{\{martin.fraenzle,paul.kroeger\}@uol.de}}
\maketitle              
\begin{abstract}
Stream-based monitoring assesses the health of safety-critical systems by transforming input streams of sensor measurements into output streams that determine a verdict.
These inputs are often treated as accurate representations of the physical state, although real sensors introduce calibration and measurement errors.
Such errors propagate through the monitor’s computations and can distort the final verdict.
Affine arithmetic with symbolic slack variables can track these errors precisely, but independent measurement noise introduces a fresh slack variable upon each measurement event, causing the monitor’s state representation to grow without bound over time.
Therefore, any bounded-memory monitoring algorithm must unify 
slack variables at runtime in a way that generates a sound approximation.

This paper introduces zonotopes as an abstract domain for online monitoring of general Robust-Lola (\rlola) specifications.
We demonstrate that zonotopes precisely capture uncertainty in the monitor's state and that their over-approximation produces a sound, bounded-memory monitor.
We present a comparison of different zonotope over-approximation strategies for runtime monitoring, evaluating their performance and false-positive rates.
Finally, we implement the most effective strategy in \rlola and demonstrate improvements in both precision and runtime compared to existing approaches based on realistic benchmarks.
\end{abstract}
\input{intro}
\input{motivation}

\input{related}
\input{prelims}
\input{approach}
\input{evaluation}
\input{conclusion}

\input{appendix}

\bibliographystyle{splncs04}
\bibliography{references}

\end{document}

%% file: commands.tex
\definecolor{bluekeywords}{rgb}{0.13, 0.13, 1}
\definecolor{greentypes}{rgb}{0, 0.5, 0}
\definecolor{inferedgreentypes}{rgb}{1.0, 0.2, 0}
\definecolor{orangecomments}{rgb}{1, 0.5, 0.1}
\definecolor{redstrings}{RGB}{171, 114, 2}
\definecolor{graynumbers}{rgb}{0.5, 0.5, 0.5}
\definecolor{goldcomments}{rgb}{0.6, 0.4, 0.08}

\makeatletter
\newcommand{\replunderscores}[1]{\expandafter\@repl@underscores#1_\relax}

\def\@repl@underscores#1_#2\relax{%
    \ifx \relax #2\relax
        #1%
    \else
        #1%
        \textunderscore
        \@repl@underscores#2\relax
    \fi
}
\makeatother

\lstdefinelanguage{Lola}{
  keywords=[0]{input, output, trigger, constant, import, spawn, eval, close, with, when},
  moredelim=**[is][\transparent{0.6}]{?}{?},
  moredelim=**[is][\color{greentypes}@]{@}{@},
  keywordstyle=[0]\bfseries\color{bluekeywords},
  keywords=[1]{if, then, else, aggregate, defaults, offset, prev, by, or, to, sin, cos, abs, hold, over, using, over_instances},
  keywords=[2]{Variable, String, Int, Int64, UInt, UInt64, Bool, Float32, Float64, Float},
  keywordstyle=[2]\color{greentypes},
  sensitive=false,
  comment=[l]{//},
  morecomment=[s]{/*}{*/},
  morestring=[b]',
  morestring=[b]",
  literate={\\@}{@}1
}
\makeatletter
\newcommand\Nat{\mathbb{N}}
\newcommand\Real{\mathbb{R}}
\newcommand\Bool{\mathbb{B}}

\newcommand{\rlola}{\textsc{RLola}\xspace}

\newcommand{\stream}[1]{{\footnotesize\texttt{\replunderscores{#1}}}\xspace}
\newcommand{\keyword}[1]{{\color{bluekeywords}\footnotesize\texttt{#1}}\xspace}
\newcommand{\lolatype}[1]{{\color{greentypes}\footnotesize\texttt{#1}}\xspace}

\newcommand{\is}{i}
\newcommand{\os}{o}
\newcommand{\cv}{\Delta}
\newcommand{\sv}{\varepsilon}

\newcommand{\Vin}{\mathbb{V}_\mathit{in}}
\newcommand{\Vout}{\mathbb{V}_\mathit{out}}
\newcommand{\Vcv}{\mathbb{V}_{\mathit{c}}}
\newcommand{\Vsv}{\mathbb{V}_{\mathit{s}}}
\newcommand{\syn}[1]{{\color{brown}\textbf{\texttt{#1}}}}
\newcommand{\synTy}[1]{{\textbf{\texttt{#1}}}}
\newcommand{\stradd}[2]{#1 \ \syn{+} \ #2}
\newcommand{\strite}[3]{\syn{if} \ #1 \ \syn{then} \ #2 \ \syn{else} \ #3}
\newcommand{\strlast}[2]{#1 \syn{.\texttt{prev}(}#2 \syn{)}}
\newcommand{\strdef}[2]{\syn{output} \ #1\ \syn{:=} \ #2}
\newcommand{\sstrdef}[1]{\syn{output} \ #1\ \synTy{:Variable}}
\newcommand{\cstrdef}[1]{\syn{constant} \ #1\ \synTy{:Variable}}
\newcommand{\smap}[1]{\mathit{Smap}(#1)}
\newcommand{\den}[1]{\llbracket #1 \rrbracket}
\newcommand{\pden}[1]{\overline{\den{#1}}{\vphantom{\den{#1}}}}
\newcommand{\rhoIn}{{\rho_{\textit{in}}}}
\newcommand{\cslacks}{\Sigma_c}
\newcommand{\sslacks}{\Sigma_s}
\newcommand{\fresh}[1]{\mathit{fresh}(#1)}

\newcommand{\abs}[1]{\lvert #1 \rvert}

%% file: intro.tex
\section{Introduction}
Stream-based runtime monitoring is a safety assurance mechanism for cyber-physical systems.
A monitor observes the system behavior and checks it against a given specification.
In stream-based specifications, observations appear as input streams that capture sensor data or other system outputs.
Output streams are defined by equations that transform and combine these inputs to assess system health.
Because many cyber-physical systems are safety-critical, the precision of the monitor’s verdicts is essential.

Monitoring frameworks often assume that input values perfectly represent the physical system.
In practice, however, sensor measurements include calibration errors and noise.
As measurements accumulate in the monitor’s state, their inaccuracies can propagate and multiply through computations, affecting the final verdict.

A simple way to model such inaccuracies is to lift stream values from scalars to intervals, as done in robust semantics for signal-temporal logic~\cite{ViscontiEA:IntervalBasedSTRELMonitoring}.
However, interval arithmetic introduces the aliasing problem: subtracting an interval from itself yields an interval double the size of the error bound (e.g., $[2,6] - [2,6] = [-4,4]$), producing overly pessimistic verdicts.

An alternative methodology for capturing measurement noise, previously explored, uses slack variables to model interval-bounded measurement errors~\cite{RobustSTL,DBLP:conf/rv/FinkbeinerFKK24}.
Slack variables symbolically represent the interval of $[-1, 1]$ and, using affine arithmetic, can also model the above interval as the affine form $x=4+2\epsilon$ for a slack variable $\epsilon$.
As slack variables are treated symbolically, the computation of $x-x$ indeed leads to the expected result of zero: $4+2\epsilon - (4+2\epsilon) = 0$.

However, if the measurement errors are independent, such as those induced by per-sample random measurement noise, an independent (fresh) slack variable is required for each measurement.
As input measurements can accumulate in stream values indefinitely, the monitor's memory requirement becomes trace-length dependent, as a new symbolic variable is memorized in each step.
Therefore, any trace-length independent runtime monitor must unify slack variables at runtime to keep its internal state from growing beyond any bound.

\rlola, an extension of Lola~\cite{Lola}, incorporates slack variables to cover inter\-val-bounded noise~\cite{DBLP:conf/rv/FinkbeinerFKK24}.
Prior work identifies a language fragment in which a bounded number of slack variables suffices for exact monitors.
This paper addresses the \emph{general case}, where arbitrary \rlola specifications may introduce an unbounded number of slack variables.

\emph{Contribution.} This paper introduces zonotopes as an abstract domain for bounded-memory monitoring of general \rlola specifications.
Zonotopes naturally capture the affine structure induced by symbolic slack variables in the monitor's memory and enable sound over-approximations when the monitor state must be compressed.
Unlike their traditional use in state estimation and reachability analysis, this paper investigates zonotopes in the context of runtime monitoring, with the primary objective of Boolean verdict precision rather than state-estimation accuracy.

We evaluate several zonotope approximation strategies in the context of runtime monitoring and analyze their impact on monitoring precision and runtime overhead.
Finally, we implement the approach in a practical \rlola monitoring framework and demonstrate substantial improvements in both precision and performance compared to existing symbolic monitoring approaches.


%% file: motivation.tex
\subsection{Motivating Example}\label{sec:motivation}
\begin{figure}
    \centering
    \includegraphics[width=0.4\linewidth]{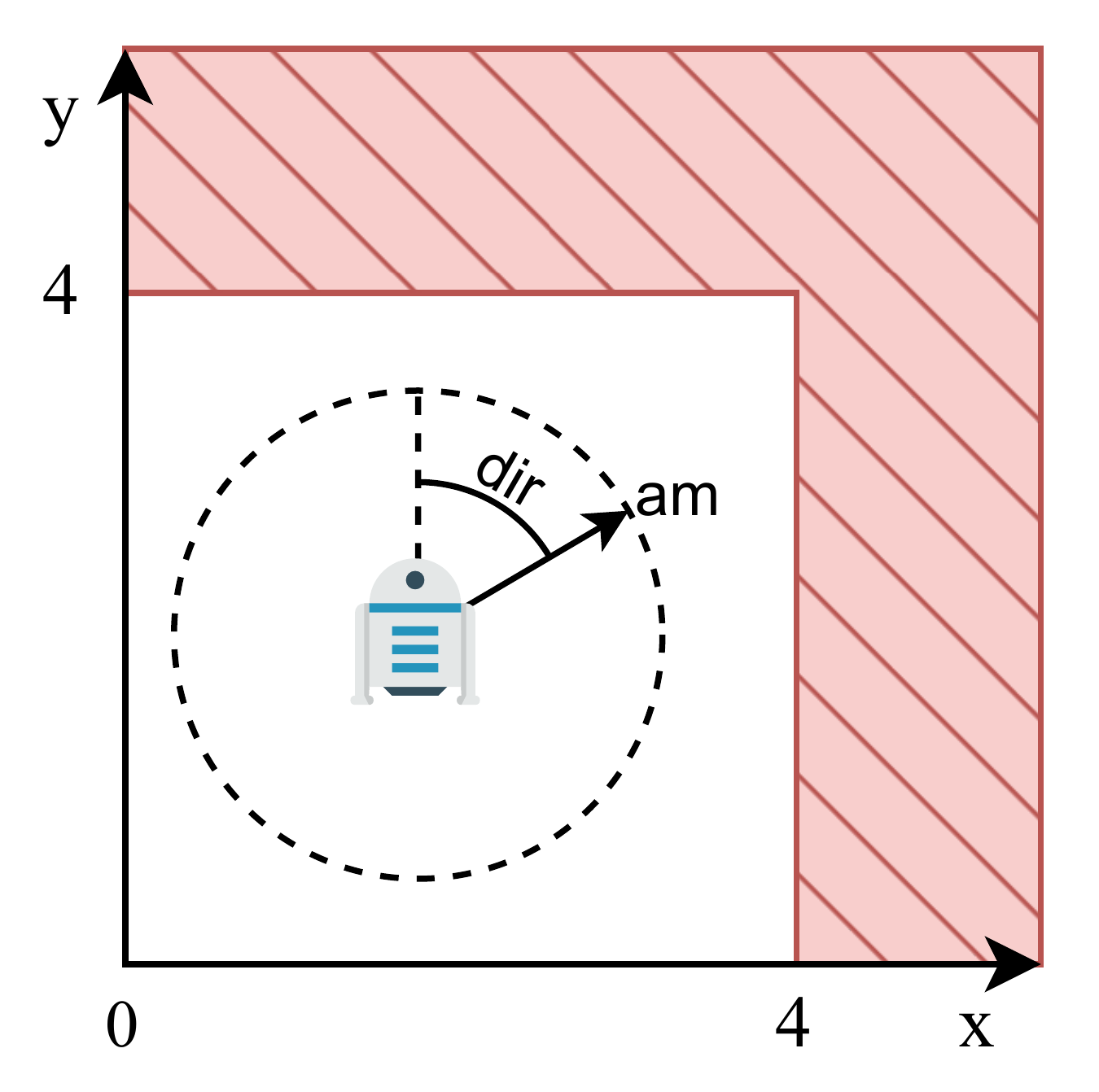}
    \caption{An industrial robot that can move in the two dimensions of a plane.}
    \label{fig:motivation}
\end{figure}

This section introduces the motivating example shown in \Cref{fig:motivation}, which is based on an industrial robot that can move omnidirectionally.
For safe operation, the robot must not enter the upper or right region of the factory where the $x$ or $y$ coordinate exceeds four.
The robot estimates its position using its measured acceleration, direction, and the elapsed time between measurements.
Monitoring whether the robot remains within its working area is a standard runtime monitoring task.
The monitor must track the robot’s position in the same way the robot does and compare it against the specified bounds.
Because acceleration measurements are noisy, the monitor must account for this noise in its computations to produce an accurate and reliable verdict.
The following specification shows how such a monitor can be defined in \rlola.
\begin{lstlisting}[label={spec:free_robot}, caption={A specification for an industrial robot that can freely move.}]
input time: Float
input dir: Float
input am: Float
output dt := time - time.offset(by: -1).defaults(to: 0.0)

constant delta: Variable
output epsilon: Variable
output a := am + 0.01 * epsilon + 0.005 * delta
output a_filter := 0.7 * a + 0.3 * a_filter.offset(by: -1).defaults(to: 0.0) 
output v := v.offset(by: -1).defaults(to: 0.0) + a_filter * dt 
output dist := 0.5 * a_filter * dt * dt + v.offset(by: -1).defaults(to: 0.0) * dt

output position_x := position_x.offset(by: -1).defaults(to: 0.0) + cos(dir) * dist
output position_y := position_y.offset(by: -1).defaults(to: 0.0) + sin(dir) * dist

trigger position_x $>_{0.01}$ 4.0 "Violated Geofence in X-Direction"
trigger position_y $>_{0.01}$ 4.0 "Violated Geofence in Y-Direction"
\end{lstlisting}
The specification begins with the \stream{time} input stream, recording the time in seconds since the start of the monitor.
The \stream{am} stream provides acceleration measurements, while the \stream{dir} input provides its direction.

Each acceleration measurement is augmented with two slack variables to capture measurement noise.
The constant slack variable \stream{delta} represents a calibration offset and is declared as a \keyword{constant} stream of type \lolatype{Variable}.
In contrast to constant variables, the output stream \stream{epsilon} of type \lolatype{Variable} introduces fresh slack variables at each step, modeling per-measurement errors.

To reduce this noise, an exponential filter is applied in \stream{a_filter}.
The filter computes a weighted sum of the current measurement and its own previous output, giving $70\%$ weight to the new sample and $30\%$ to the accumulated history. 
The \keyword{prev} operator yields the previous stream value or a default if none exists.
The position coordinates are then reconstructed by adding the distance traveled in the current step to the last known coordinate relative to the direction.

Finally, trigger streams compare the reconstructed positions to their thresholds.
Because the position streams represent value ranges, direct comparisons can yield inconclusive verdicts.
To circumvent this, \rlola extends comparison operators with an overlap percentage that specifies how to evaluate ranges against thresholds.
In this example, the range must overlap the threshold of four with at least one percent of its total width for the expression to evaluate to true.

\begin{figure}[bt]
    \centering
    \includegraphics[width=0.65\linewidth]{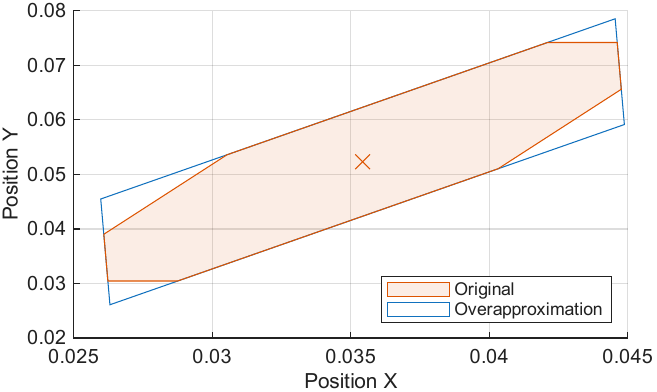}
    \vspace{-3mm}
    \caption{A plot of the zonotope spanned by the monitor state at time 4s.}
    \label{fig:zonotope_ex}
\end{figure}

Designing a general online monitoring algorithm for \rlola is non-trivial, as shown by inspecting the monitor’s memory over time.
Consider the monitor state in \Cref{tab:monitor_mem} which depicts the stream values for the \stream{position_x} and \stream{position_y} streams.
\begin{table}[b]
    \centering
    \setlength{\tabcolsep}{8pt}
    \renewcommand{\arraystretch}{1.2}
    \centering
    \begin{tabular}{l r r}
       \toprule
       \textbf{Time}  & \textbf{\stream{position_x}} & \textbf{\stream{position_y}}\\
       \midrule
       \textbf{3.1s}  & {$\!\begin{aligned}
           0.0354 + 0.0058 \epsilon_0 - 0.0001 \epsilon_1\\
           + -0.0013 \epsilon_2 + 0.0022 \Delta
       \end{aligned}$} &  {$\!\begin{aligned}
           0.0523 + 0.0103 \epsilon_0 + 0.0043 \epsilon_1\\
           + 0.0073 \Delta
       \end{aligned}$} \\
       \bottomrule
    \end{tabular}
    \caption{The monitor's memory depicted over an sample trace.}
    \label{tab:monitor_mem}
\end{table}
Slack variables produced by the \stream{epsilon} stream accumulate in these streams, causing their representations to grow with the trace length.
A trace-length-independent monitor must therefore prune slack variables at runtime.
To this end, we observe that the affine forms representing the stream values at each step form a convex set describing the monitor’s state.
This set is a zonotope, which is generally defined by a center vector and a set of column vectors called its generators.
It is then constructed as the Minkowski sum of its generators.
Here, the center contains the absolute parts of the stream values, and the generators are given by the coefficients of the slack variables.
The zonotope at time $3.1$ has the center vector and generators plotted in red in \Cref{fig:zonotope_ex}.:
\[
c = \begin{pmatrix}
0.0354\\
0.0523
\end{pmatrix}
\quad
G=\begin{pmatrix}
0.0058 & -0.0001 & -0.0013 & 0.0022\\
0.0103 &  0.0043 &  0.0    & 0.0073 
\end{pmatrix}
\]
Pruning slack variables corresponds to reducing the number of zonotope generators.
Prior work shows this reduction is exact for a fragment of \rlola specifications~\cite{DBLP:conf/rv/FinkbeinerFKK24} in which all generators are collinear.
In this example, the sine and cosine operators produce non-collinear generators, so the monitor must over-approximate the zonotope to reduce its generators, as shown in blue in \Cref{fig:zonotope_ex}.

%% file: related.tex
\subsection{Related Work}

\emph{Monitoring Logics.}
Runtime monitoring has been studied for formalisms such as LTL~\cite{LTLMonitoring}, STL~\cite{STLMonitoring}, and MTL~\cite{DBLP:conf/rv/HoOW14}. 
Stream-based languages including Lola~\cite{Lola}, RTLola~\cite{RTLola,DBLP:conf/cav/BaumeisterFKLMST24}, Striver~\cite{Striver}, and Tessla~\cite{TeSSLa} describe temporal behavior through equations over streams, but mostly do not account for measurement uncertainty.

\emph{Monitoring under Uncertainties.}
Previous work has studied monitoring under unreliable communication~\cite{DBLP:journals/sttt/KauffmanHF21}, imprecise timestamps~\cite{DBLP:conf/ifm/FranzleGLZ24}, and noisy measurements~\cite{DonzeMaler10,DBLP:conf/cav/DonzeFM13,ViscontiEA:IntervalBasedSTRELMonitoring,STLMonitoring,DBLP:conf/rv/LeuckerSS0T19,DBLP:conf/rv/FinkbeinerFKK24}. 
Most approaches assume independent interval-bounded errors~\cite{DonzeMaler10,DBLP:conf/cav/DonzeFM13,ViscontiEA:IntervalBasedSTRELMonitoring,DBLP:conf/rv/LeuckerSS0T19}. 
The ISO~EC~5725~\cite{ISO:5725} error model additionally distinguishes constant calibration error from time-varying noise and has been shown to improve precision~\cite{STLMonitoring}, but is rarely supported in monitoring~\cite{STLMonitoring,DBLP:conf/rv/FinkbeinerFKK24,DBLP:conf/ifm/FranzleGLZ24}.

\emph{Stream-based Monitoring under Uncertainties.}
Prior stream-based monitoring work addresses missing or shifted inputs~\cite{DBLP:conf/rv/LeuckerSS0T19}. 
\rlola, as introduced in~ \cite{DBLP:conf/rv/FinkbeinerFKK24} assumes all inputs being present but noisy.
That work considers only a fragment of \rlola with constant-size monitoring state, whereas this paper considers general specifications.
Kallwies et al.~\cite{DBLP:conf/atva/KallwiesLS22,DBLP:conf/cav/HiplerKLS24,DBLP:journals/infsof/HiplerKLMSW26} model uncertainty symbolically using SMT solving and interval analysis, which can lead to conservative approximations and a higher runtime. 
In contrast, our approach first computes an exact symbolic affine representation and then over-approximates it using zonotopes.

\emph{Approximating Zonotopes.}
Zonotopes are widely used in reachability analysis~\cite{DBLP:conf/cdc/AlamoB003,DBLP:conf/cdc/AlthoffF16}. 
Compared to reachability analysis, runtime monitoring continuously incorporates new measurements that refine the current state. 
State-bounded observers~\cite{DBLP:conf/cdc/Combastel05} are similar but are not designed to produce boolean state verdicts.
These works provide approximations applicable to our setting~\cite{DBLP:conf/cdc/KopetzkiSA17,DBLP:conf/cdc/Combastel05}.

%% file: prelims.tex
\section{Preliminaries}\label{sec:prelims}
In this section, we recap the syntax and semantics of \rlola as introduced in~\cite{DBLP:conf/rv/FinkbeinerFKK24}.
We restate the measurement-noise model used in the examples and explain how interval-bounded values are compared with trigger thresholds. 

\subsection{\rlola}
An \rlola specification is a set of stream equations over input streams, output streams, and slack-variable streams.
Formally, this set is defined as follows:
\begin{definition}[\rlola Syntax]\label{def:syntax}\\
Let $\mathbb{V} \triangleq \Vin \uplus \Vout \uplus \Vcv \uplus \Vsv$\\

\begin{tabular}{ll}
  \textit{Variables}         & $x \ ::= \ \is \in \Vin\ \mid\ \os \in \Vout\ \mid\ \cv \in \Vcv\ \mid\ \sv \in \Vsv$\\
  \textit{Values}             & $v \in \mathbb{R} \uplus \mathbb{B}$\\
  \textit{Stream Expressions} & $e \ ::= \ v \mid x \mid \strlast{x}{e} \mid \stradd{e_1}{e_2}$\\
                              & $\qquad\quad\mid \strite{e_b}{e_c}{e_a} \mid \ldots$\\
  \textit{Equations}          & $eq ::= \strdef{\os}{e} \mid \cstrdef{\cv}$\\
                              & $\qquad\quad\mid \sstrdef{\sv}$\\
  \textit{Specifications}     & $S ::= \emptyset \mid eq \cdot S$
\end{tabular}
\end{definition}
We do not include standard comparison operators (e.g., less, greater, equal) in stream expressions, although they can be extended by further arithmetic operations and boolean connectives.
Including such operators would enable booleanizing noisy stream values, which introduces nondeterminism when evaluating them as conditions of an if-clause.
We instead introduce dedicated comparison operators that handle noisy stream values.

The equations relate input values and slack-variable values to output values.
A model of a specification assigns a value to each variable at each time point such that all equations are satisfied.
In contrast to Lola, a specification may have multiple models for the same input stream 
due to undetermined slack-variable values.
We give the semantics of \rlola as an adaptation of the asynchronous Lola semantics~\cite{DBLP:journals/corr/abs-2509-06724} and the original \rlola semantics~\cite{DBLP:conf/rv/FinkbeinerFKK24}.
A stream is defined as a discrete sequence of values and variables are interpreted by streams:
\[
    \mathit{Stream} \triangleq \Real^\Nat\ \cup\ \Bool^\Nat\hspace{1cm}
    \smap{\mathbb{V}} \triangleq \mathit{Stream}^\mathbb{V}
\]
Expression are given semantics relative to a model $\rho \in \smap{\mathbb{V}}$ at time $n \in \mathbb{N}$:
\begin{definition}[Semantics of Stream Expressions]\label{def:sem-expr}
Let $v \in \Real \cup \Bool$ and $x \in \mathbb{V}$
\upshape
\begin{align*}
\den{v}^{n}_\rho &\triangleq v\\
\den{x}^{n}_\rho &\triangleq \rho(x)(n)\\
\den{\stradd{e_1}{e_2}}^{n}_\rho &\triangleq \den{e_1}^{n}_\rho + \den{e_2}^{n}_\rho\\
\den{\strlast{x}{e}}^{n}_\rho &\triangleq \begin{cases}
    \begin{tabular}{ll}
      $\rho(x)(n-1)\quad$ &if $n > 0$\\
      $\den{e}^{n}_\rho\quad$ &otherwise
    \end{tabular}
\end{cases}\\
\den{\strite{e_b}{e_c}{e_a}}^{n}_\rho &\triangleq \begin{cases}
    \begin{tabular}{ll}
      $\den{e_c}^{n}_\rho\quad$ &if $\den{e_b}^{n}_\rho = \top$ \\
      $\den{e_a}^{n}_\rho\quad$ &otherwise
    \end{tabular}
\end{cases}
\end{align*}
\end{definition}
A stream equation denotes the set of all models that satisfy it, and a specification denotes the intersection of the sets given by its equations.
\begin{definition}[Semantics of Specifications]\label{def:sem-spec}
\upshape
  \begin{align*}
    \den{\strdef{\os}{e}} &\triangleq \{ \ \rho \in \smap{\mathbb{V}} \mid \forall n \in \mathbb{N}. \rho(\os)(n) = \den{e}^n_\rho \ \}\\
    \den{\cstrdef{\cv}} &\triangleq \{ \ \rho \in \smap{\mathbb{V}}  \mid \exists \delta \in [-1,1]. \forall n \in \mathbb{N}.\ \rho(\cv)(n) = \delta \ \}\\
    \den{\sstrdef{\sv}} &\triangleq \{ \ \rho \in \smap{\mathbb{V}}  \mid \forall n \in \mathbb{N}.\ \rho(\sv)(n) \in [-1,1] \ \}\\
    \den{ \emptyset } &\triangleq \smap{\mathbb{V}}\\
    \den{ \mathit{eq} \cdot S } &\triangleq \den{\mathit{eq}} \cap \den{S}
  \end{align*}
\end{definition}
The standard well-formedness conditions from Lola~\cite{Lola} also apply to \rlola.
Given a sequence of input measurements $\rhoIn \in \smap{\Vin}$, we define the set of valid models for a specification $S$ on these input measurements as:
\[
    \den{S}_\rhoIn \triangleq \{ \rho \in \den{S}  \mid \rho\downarrow_{\Vin} = \rhoIn \}
\]
where $\rho\downarrow_{\Vin}$ is the projection of $\rho$ on the input variables.

A monitor for an \rlola specification computes a symbolic representation of $\den{S}_{\rhoIn}$.
The main question for robust semantics is how to evaluate boolean conditions over such sets of possible states. 

\paragraph{Trigger Conditions in \rlola.}
We use the ternary predicates $>_p$ and $<_p$ to compare a value range with a scalar threshold using an overlap percentage.
For a stream $x \in \mathbb{V}$, the range of possible values at time $n$ under measurements $\rhoIn$ is defined as
$
    \den{x}_\rhoIn^n \triangleq \{\ v \mid \exists \rho \in \den{S}_\rhoIn \text{ with } \rho(x)(n) = v \ \}
$.
For $p \in [0,1]$ and $v \in \mathbb{R}$, the boolean value of the predicates is
\begin{align*}
    \den{x >_p v}_\rhoIn^n &\triangleq (\mathit{max}(\den{x}_\rhoIn^n) - v)/(\mathit{max}(\den{x}_\rhoIn^n) - \mathit{min}(\den{x}_\rhoIn^n)) > p\,,\\
    \den{x <_p v}_\rhoIn^n &\triangleq (v - \mathit{min}(\den{x}_\rhoIn^n))/(\mathit{max}(\den{x}_\rhoIn^n) - \mathit{min}(\den{x}_\rhoIn^n)) > p\,.
\end{align*}

These predicates allow specifications to express verdicts beyond the usual inconclusive case when intervals overlap the threshold.
The overlap percentage determines what fraction of the interval must lie above (or below) the threshold. 

\subsection{Error Model}
As in~\cite{RobustSTL,DBLP:conf/rv/FinkbeinerFKK24}, the examples use the error model defined in ISO~5725~\cite{ISO:5725}, which decomposes measurement noise into a constant calibration error and a random per-measurement error.
This model can be encoded in \rlola using constant slack variables and slack-variable streams, as demonstrated in \Cref{sec:motivation}.

We restate the definition of consistency from~\cite{RobustSTL}, which characterizes the ground-truth values consistent with a given measurement series:
\begin{definition}[Consistency]\label{def:consistency}
Let $\delta \ge 0$ be the maximal sensor calibration offset and  $\mu \ge 0$ be the maximal random measurement error. Let $\tau$ be the ground-truth time series of a physical property. Then the input stream $m \in \mathit{Stream}$ is a possible time series over $\tau$ of sensor measurements at time points $t$ iff: $\exists \Delta \in [-\delta,\delta].\ \forall t \in \Nat.\ \exists \varepsilon \in [-\mu,\mu].\ \tau(t) + \varepsilon + \Delta = m(t).$
\end{definition}

%% file: approach.tex
\section{Approximate Online Monitoring of \rlola}\label{sec:approach}
This section presents a novel 
online monitoring algorithm for \rlola.
A major challenge is ensuring bounded memory, as an online monitor must run indefinitely alongside the system.
Semantically, a monitor computes the set $\den{S}_\rhoIn$ for the measurements $\rhoIn$, and trigger conditions are evaluated over this set using the 
predicates introduced above.
Efficiently representing this set is difficult because it is uncountable in general, since slack variables range over the reals.
The algorithm addresses this by constructing a symbolic representation of the state space, where slack variables are encoded as symbolic variables.


\subsection{Symbolic Semantics}
We assume two disjoint infinite sets of symbolic variables: $\cslacks$ for constant slack variables and $\sslacks$ for per-measurement slack variables.
The function $\fresh{\Sigma}$ returns a previously unused symbolic variable from $\Sigma$.
The symbolic domain is defined as affine forms~\cite{AffineArithmetic} over variables in $\cslacks$ and $\sslacks$:
\[
    \mathit{Affine}_\Sigma \triangleq c + \sum_{\varepsilon \in \Sigma} \mu_\varepsilon * \varepsilon
\]
where $c \in \Real$ and $\mu \in \Real^\Sigma$.
The right-hand side of the equation is purely syntactic:
The arithmetic operators are only ever evaluated once an interpretation for the symbolic variables is given.
Affine forms symbolically represent an interval in $\Real$, as 
symbolic variables 
range over the interval $[-1,1]$.
Arithmetic operations applied to affine forms follow the standard rules for affine arithmetic~\cite{AffineArithmetic}.


The symbolic semantics for \rlola tracks slack variables symbolically by extending the stream value domain to affine forms:
$\mathit{Stream} \triangleq \mathit{Affine}^\Nat \cup\ \Bool^\Nat$.
It is equivalent to the relational semantics except that constant slack variables and slack-variable streams evaluate to symbolic variables.
Concretizations of symbolic states instantiate these variables.
\begin{definition}[Symbolic Semantics of Specifications]\label{def:sym-sem-spec}
\upshape
  \begin{align*}
    \pden{\strdef{\os}{e}} &\triangleq \{ \ \rho  \in \smap{\mathbb{V}} \mid \forall n \in \mathbb{N}. \rho(\os)(n) = \den{e}^n_\rho \ \}\\
    \pden{\cstrdef{\cv}} &\triangleq \{ \ \rho  \in \smap{\mathbb{V}} \mid \\
    &\qquad\ \delta = \fresh{\cslacks} \land \forall n \in \mathbb{N}.\ \rho(\cv)(n) = \delta \ \}\\
    \pden{\sstrdef{\sv}} &\triangleq \{ \ \rho  \in \smap{\mathbb{V}} \mid \forall n \in \mathbb{N}.\ \rho(\sv)(n) = \fresh{\sslacks} \ \}\\
    \pden{ \emptyset } &\triangleq \smap{\mathbb{V}}\\
    \pden{ \mathit{eq} \cdot S } &\triangleq \pden{\mathit{eq}} \cap \pden{S}
  \end{align*}
\end{definition}
The semantics for stream expressions stay unchanged, but use affine arithmetic for all operations.
Symbolic models are considered equivalent up to consistent renaming of symbolic variables.
Given an assignment to symbolic variables, affine forms evaluate as follows:
Let $\Gamma \triangleq \Sigma \rightarrow [-1, 1]$ be an assignment to slack variables, then:
\[
\den{c + \sum_{\varepsilon \in \Sigma} \mu_\varepsilon * \varepsilon}_\Gamma \triangleq c + \sum_{\varepsilon \in \Sigma} \mu_\varepsilon * \Gamma(\varepsilon)
\]
Using this, we define the concretization of both, a symbolic model $\rho_s \in \pden{S}$ to a set of concrete models $\rho \in \den{S}$ and the concretization of all symbolic models in the specification that agree with some $\rhoIn$.
\begin{align}
    \abs{\rho_s} &\triangleq \{\ \rho \mid \exists \Gamma. \forall x \in \mathbb{V}. \forall n \in \Nat. \rho(x)(n) = \den{\rho_s(x)(n)}_\Gamma \ \}\label{eq:concrete_model}\\
    \abs{\pden{S}_\rhoIn} &\triangleq \bigcup\nolimits_{\rho_s \in \pden{S}_\rhoIn} \abs{\rho_s}\label{eq:concrete_spec}
\end{align}
%
%
%
Based on these concretizations, we state the correctness lemma for the symbolic semantics. 
\begin{lemma}[Symbolic Correctness]
    $\forall \rhoIn. \abs{\pden{S}_\rhoIn} = \den{S}_\rhoIn$
\end{lemma}
Before addressing the correctness of the symbolic semantics, we highlight two sources of non-determinism in \rlola specifications that yield multiple models in the non-symbolic semantics.
First, the non-deterministic choice of slack variable values leads to multiple models consistent with a measurement series $\rhoIn$.
Second, a specification may contain self-referential equations such as $\strdef{x}{x}$, which are satisfied by any value of $x$ and thus introduce further non-determinism.
As in Lola~\cite{Lola}, such equations are excluded by a well-formedness criterion to ensure determinism up to measurement noise.

The correctness lemma follows from how the symbolic semantics resolves the specification’s non-determinism during concretization.
First, the existential quantification in \Cref{eq:concrete_model} fixes the measurement noise by assigning values to symbolic variables.
In the symbolic semantics, constant slack definitions yield the same symbolic variable at all time steps, and slack-variable streams introduce a fresh symbolic variable per step, matching the quantifier behavior in the non-symbolic semantics.
Second, \Cref{eq:concrete_spec} resolves the non-determinism of self-referential streams by taking the union over the concretizations of all symbolic models.


\subsection{Symbolic Algorithm}
We present an algorithm that computes a symbolic model for a specification $S$ and input measurements $\rhoIn$.
We assume that $S$ is well-formed, so $\pden{S}\rhoIn$ contains exactly one model.
The algorithm follows the structure of the Lola monitoring algorithm~\cite{Lola} and maintains two stores: the resolved store $R$, which contains equations of the form $x_i = c + \sum_{\varepsilon \in \Sigma} \mu_\varepsilon \cdot \varepsilon$, and the unresolved store $U$, which contains equations of the form $x_i = e_x$.
The algorithm is given below.

\paragraph{Symbolic Evaluation Algorithm.}
Initially, let $U$ and $R$ be the empty set, and 
$C$ be a map from constant slack variables to symbolic variables defined as follows:
\[
    C = \{\ (\cv, \fresh{\cslacks}) \mid \cv \in \Vcv\ \}
\]
For each input from $\rhoIn$ that becomes available at time $0 \leq t \leq \Nat$, the following steps are performed:
\begin{enumerate}
    \item Record the input streams as resolved: $R = R \cup \{\ \is_t = \rhoIn(\is)(t) \mid \is \in \Vin\ \}$
    \item Resolve the constant slack variables: $R = R \cup \{\ \cv_t = C(\cv) \mid \cv \in \Vcv\ \}$
    \item Draw symbolic variables for slack-variable streams: 
        \[R = R \cup \{\ \sv_t = \fresh{\sslacks} \mid \sv \in \Vsv\ \}\]
    \item Add the defining equations for output streams to the unresolved store:
        \[U = U \cup \{\ \os_t = e \mid \strdef{\os}{e} \in S\ \}\]
    \item Simplify the equations in $U$ as much as possible using the equations in $U \cup R$.
        If an equation becomes of the form $\os_t = c + \sum_{\varepsilon \in \Sigma} \mu_\varepsilon * \varepsilon$ remove it from $U$ and add it to $R$.
    \item Evaluate trigger conditions based on $R$ and report the violations.
    \item Prune stale equations from $R$: Remove equations $x_j$ for $x \in \mathbb{V}$ and $j < t-1$.
\end{enumerate}
The expressions of the equations in $U$ at time $t$ are rewritten according to the following rules:
\begin{itemize}
    \item Replace $x \in \mathbb{V}$ with $x_i$
    \item Rewrite $\stradd{e_1}{e_2}$ using affine arithmetic, if $e_1$ and $e_2$ are fully resolved.
    \item Rewrite $\strlast{x}{e}$ with $x_{i-1}$ if $t > 0$ or with $e$ otherwise.
    \item Rewrite $\strite{e_b}{e_c}{e_a}$ to $e_c$ if $e_b = \top$ or else to $e_a$.
\end{itemize}
The algorithm computes symbolic stream values step-by-step, using only values from the previous 
step, as 
\rlola 
streams may only reference values from the immediately preceding step.
Older equations are removed from $R$ in step 7.
Steps 1--4 correctly implement the semantics of equations, and correctness follows because the rewriting rules in step 5 match the expression semantics.

\subsection{Zonotope Approximation}
Although the algorithm represents $\den{S}_\rhoIn$ exactly, it may require unbounded memory: the number of symbolic variables in the affine forms in $R$ can grow without limit as demonstrated in \Cref{sec:motivation}.
To address this, we approximate $R$ after each iteration using zonotope over-approximations.

As described in previous work about affine arithmetic~\cite{AffineArithmetic}, the joint domain of multiple affine forms constitutes a zonotope.
Hence, all affine forms in the resolved set $R$ form a zonotope that represents the current monitoring state.
A zonotope $Z(c,G)$ is a symmetric convex set with center $c \in \Real^d$ and generator matrix $G \in \Real^{d \times k}$ of dimension $d$ with $k$ spanning vectors:
\[
Z(c, G) \triangleq \{\ c + G \cdot \overline{\varepsilon} \mid \overline{\varepsilon} \in [-1,1]^d \ \}
\]
A two-dimensional example is shown in~\Cref{fig:zonotope_ex}.
In our setting, the dimension of the zonotope equals the number of equations, and the number of generators equals the number of symbolic variables in $R$.

To align equations in $R$ with rows of a zonotope, we fix an arbitrary order $\leq$ over stream variables $\mathbb{V}$ and list them as $x_1,\dots,x_d$.
A bijection $\alpha$ from lists of affine forms to zonotopes of the same dimension is then defined by mapping each affine form to its center and coefficient vector \cite{AffineArithmetic}:
\[
    \alpha(R) = Z(c,G) \qquad c_i = \hat{c}(R(x_i)) \qquad G_{i,\cdot} = \hat{\mu}(R(x_i))
\]
Where $i$ is a row index and $\hat{c}(\cdot)$ and $\hat{\mu}(\cdot)$ return the center and coefficients of the affine form.
%
The inverse $\alpha^{-1} = \beta$ retrieves the affine forms in the expected way.

We extend the symbolic evaluation algorithm with step eight, which approximates $R$ via zonotopes.
Let $\mathit{size}(Z(c,G))$ be the number of generators, i.e., the number of non-zero columns of $G$. We add:
\begin{enumerate}
  \setcounter{enumi}{7}
  \item If $\mathit{size(\alpha(R))} > k$, over-approximate $R$ as $R^{k} = \beta(A^k(\alpha(R)))$
\end{enumerate}
Here $k$ is the bound on the number of symbolic variables, and $A^k$ is a function that over-approximates a zonotope with one having at most $k$ generators but the same center.

We reason about the soundness and completeness of the algorithm with respect to trigger conditions. 
Ternary predicates are evaluated 
over $R$:
\begin{align*}
    \pden{x >_p v}_R &\triangleq (\mathit{max}(R(x)) - v)/(\mathit{max}(R(x)) - \mathit{min}(R(x)) > p
\end{align*}
    
\begin{theorem}[Correctness]\label{theo:alg_correctness}
Let $\rhoIn$ be a series of measurements and 
$R$ be the resolved set computed by the evaluation algorithm at time $t$ for $\rhoIn$. It holds that:
\begin{align*}
    p \leq 0.5 \land \den{x >_p v}_\rhoIn^t &\implies \pden{x >_p v}_R & p\leq 0.5 \land \den{x <_p v}_\rhoIn^t &\implies \pden{x <_p v}_R\\
    p \geq 0.5 \land \pden{x >_p v}_\rhoIn^t &\implies \den{x >_p v}_R & p \geq 0.5 \land \pden{x <_p v)}_\rhoIn^t &\implies \den{x <_p v}_R
\end{align*}
\end{theorem}
To prove this theorem, let $u$ and $l$ be the upper and lower bounds for $x$ computed by the maximum and minimum operators when evaluating $\den{x >_p v}_\rhoIn^t$. Let $\hat{u}$ and $\hat{l}$ be the bounds for $\pden{x >_p v}_R$.
Assume the over-approximation of $R$ widened these bounds by a value $\eta > 0$.
Given the correctness of the algorithm and the fact that zonotopes are symmetric around their center and the approximation method preserves the center, it follows that $\hat{l} = l - \eta$ and $\hat{u} = u + \eta$.
For $p \in [0, 0.5]$ follows $(u-v)/(u-l) > p \Rightarrow (\hat{u} - v)/(\hat{u} - \hat{l}) > p$.
For $p \in [0.5, 1]$ follows $(u-v)/(u-l) > p \Leftarrow (\hat{u} - v)/(\hat{u} - \hat{l}) > p$.

\Cref{theo:alg_correctness} means that the algorithm is sound for any $p \leq 0.5$, i.e., that every violation reported according to the non-symbolic semantics, are also reported by the algorithm.
For $p \geq 0.5$, the algorithm is complete, i.e., every violation reported by the algorithm is indeed a violation according to the semantics.
For $p = 0.5$ the algorithm is sound and complete.
These properties are direct consequences of the point-wise symmetry of the over-approximation.

%% file: evaluation.tex
\section{Evaluation and Implementation}\label{sec:eval}
This section first evaluates existing zonotope approximation strategies in the context of runtime monitoring.
Based on this analysis, the most effective method is integrated into the \rlola monitoring framework.
The resulting tool is then evaluated on realistic benchmarks and compared against existing approaches with respect to precision and runtime.
The evaluation addresses the following research questions:
\begin{itemize}
    \item []\textbf{RQ1:} Which zonotope approximation methods are best suited for runtime monitoring?
    \item []\textbf{RQ2:} How do the performance and precision of the proposed tool compare to the state of the art?
    \item []\textbf{RQ3:}  What level of expressivity does the approach provide, and does it suffice for realistic monitoring applications?
\end{itemize}
As a baseline, we use the SMT-based Lola extension for symbolic inputs~\cite{DBLP:journals/infsof/HiplerKLMSW26} (UnaaLola), modeling slack variables as uncertain inputs over $[-1,1]$ and constant slack variables through equality assumptions.

\subsection{Approximation Methods}
We evaluate the approximation methods using the omni-directional robot example from \Cref{sec:motivation}.
Using randomly sampled traces generated from this system, we measure both the loss in precision between exact and approximated zonotopes and the resulting false-positive rates.

To quantify precision loss, we compare the interval hulls of zonotopes using the mean squared error between corresponding dimensions.
Unlike volume-based or Euclidean Hausdorff metrics~\cite{DBLP:conf/cdc/KopetzkiSA17,11028940}, this metric directly reflects trigger evaluation, which depends on axis-aligned bounds.
Benchmark traces are generated from random-walk trajectories with additional measurement noise according to the error model.
Resolved sets are computed using a MATLAB implementation, while zonotope approximations are obtained with the CORA toolbox~\cite{Althoff2015ARCH}.
We evaluate all sound fixed-size approximation methods provided by the toolbox: combastel~\cite{DBLP:conf/cdc/Combastel05}, girard~\cite{DBLP:conf/hybrid/Girard05}, methA~\cite{dissAlthoff}, scott~\cite{DBLP:journals/automatica/ScottRMB16}, pca~\cite{DBLP:conf/cdc/KopetzkiSA17}, and the interval hull.

\begin{figure*}[ht]
    \centering
    \begin{subfigure}{\linewidth}
        \setlength{\tabcolsep}{8pt}
        \renewcommand{\arraystretch}{1.2}
        \centering
        \begin{tabular}{l r r r r r r r}
           \toprule
           \textbf{Method}  & \textbf{Girard} &  \textbf{PCA} & \textbf{Scott} & \textbf{MethA} & \textbf{Combastel}  & \textbf{Box} \\
           \midrule
           \textbf{FPR}
             & $0.0254$
             & $0.0265$
             & $0.0265$
             & $0.0387$
             & $0.0424$
             & $0.0657$\\
           \bottomrule
        \end{tabular}
        \caption{False-positive rates of the approximation methods.}
        \label{tab:false_positives}
    \end{subfigure}
    \begin{subfigure}{0.45\linewidth}
        \includegraphics[width=\textwidth]{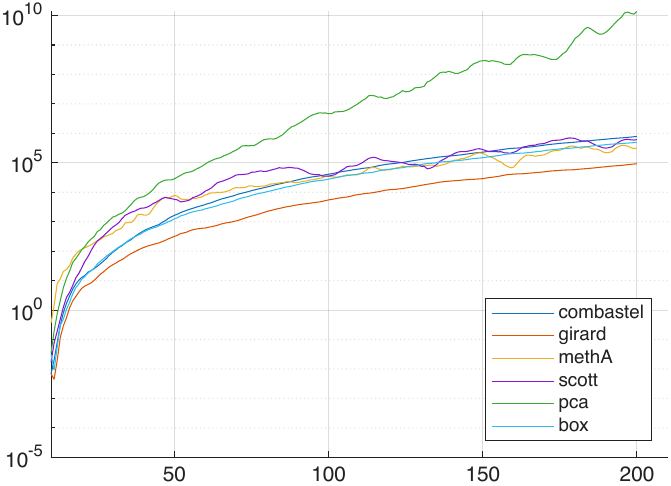}
        \caption{Mean approximation error over time (logarithmic scale).}
        \label{fig:eval_adapted:time}
    \end{subfigure}
    \hfill
    \begin{subfigure}{0.45\linewidth}
        \includegraphics[width=\textwidth]{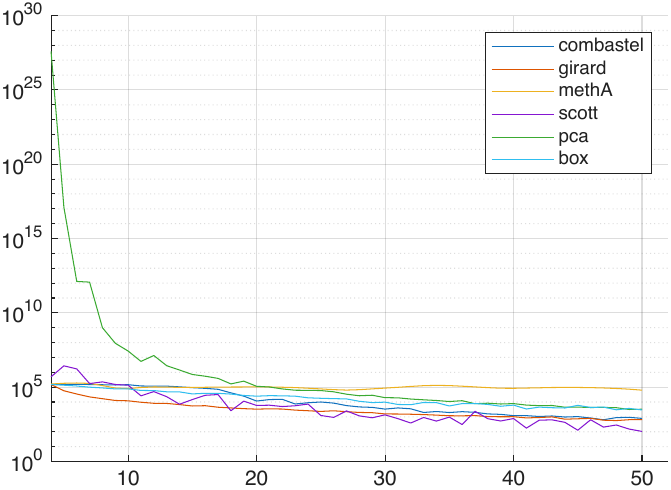}
        \caption{Mean approximation error for varying slack-variable limits.}
        \label{fig:eval_adapted:limit}
    \end{subfigure}
    \caption{Approximation error and false-positive rates for \Cref{spec:free_robot}.}
\end{figure*}
We first analyze approximation error over time and for different slack-variable limits.
\Cref{fig:eval_adapted:time} shows the average approximation error over traces of length 200 with a total slack-variable limit of $k=8$.
For all methods, the error increases over time, with \emph{pca} exhibiting the fastest degradation.
\Cref{fig:eval_adapted:limit} shows the average approximation error for different slack-variable limits.
Increasing the limit generally improves precision, although the magnitude of the improvement differs across methods.
These results indicate that, with the exception of \emph{methA}, retaining additional slack variables improves the precision of monitoring verdicts.
The observed differences are reflected in the false-positive rates of trigger conditions shown in Table~\ref{tab:false_positives}, averaged over ten traces of length 1000.

Based on this, we answer \textbf{RQ1} and implemented the Scott~\cite{DBLP:journals/automatica/ScottRMB16} and Girard~\cite{DBLP:conf/hybrid/Girard05} approximation methods.

\subsection{The \rlola Monitoring Framework}
To answer the remaining two research questions, we implemented the approach in a Rust-based \rlola monitoring framework supporting arbitrary \rlola specifications~\footnote{The implementation and evaluation is available at: \url{https://github.com/reactive-systems/Robust-Lola}}. 
The framework provides a native implementation of symbolic slack variables, precise pruning for the bounded-memory fragment~\cite{DBLP:conf/rv/FinkbeinerFKK24} (denoted collinear), and zonotopic approximations based on the online monitoring algorithm from~\Cref{sec:approach}.

\begin{figure}[H]
    \centering
    \begin{subfigure}{0.45\linewidth}
        \includegraphics[width=\textwidth]{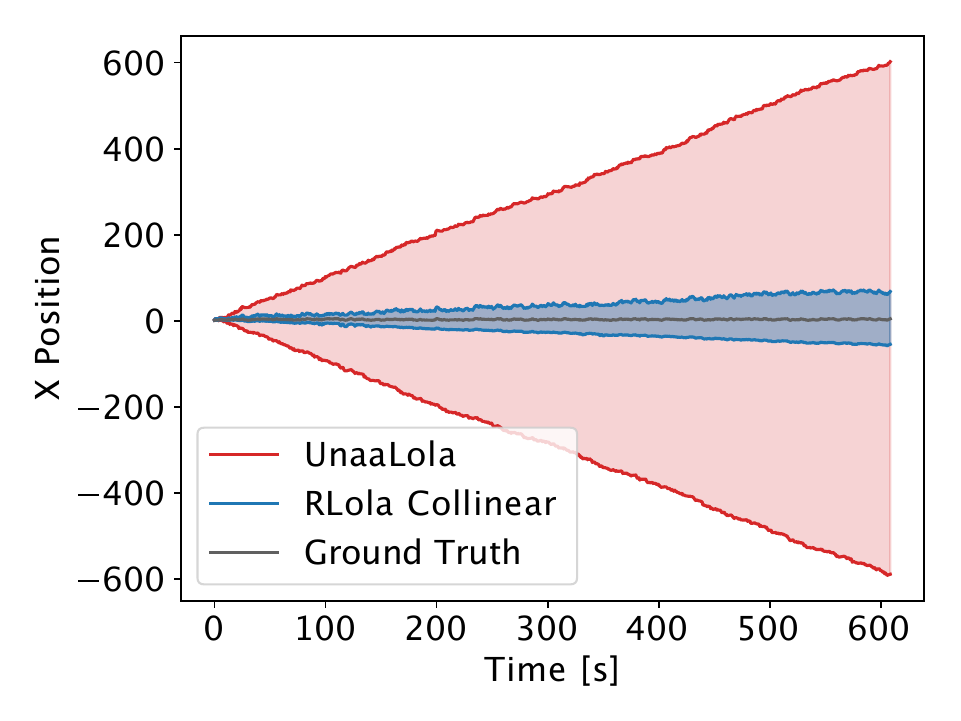}
        \caption{Uncertainty range around the robot's $x$ position.}
        \label{fig:discrete:value}
    \end{subfigure}
    \hfill
    \begin{subfigure}{0.45\linewidth}
        \includegraphics[width=\textwidth]{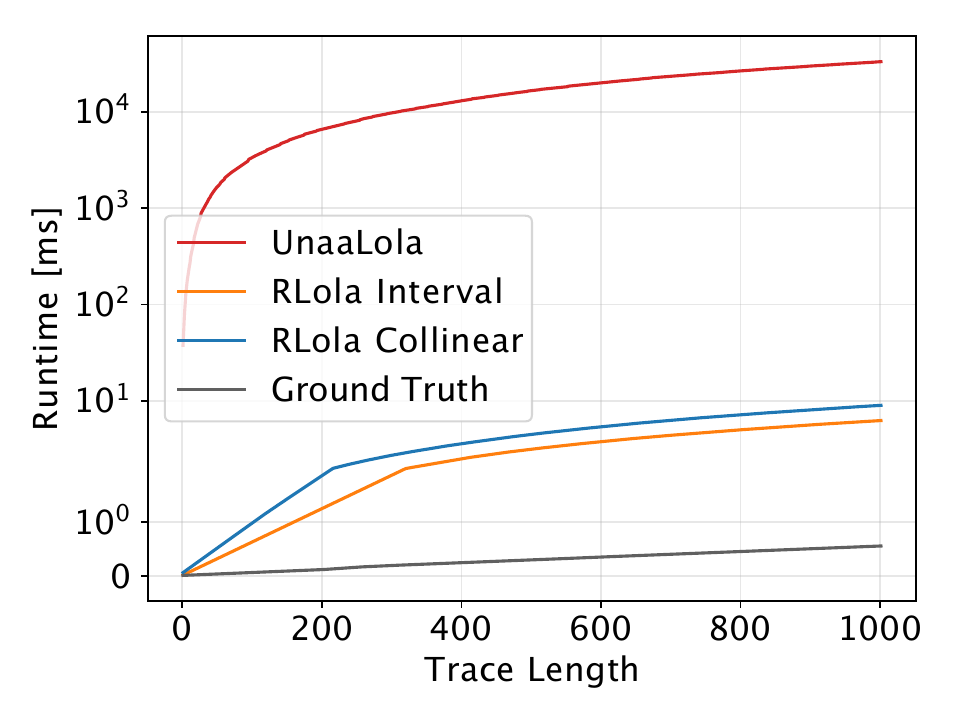}
        \caption{Runtime for increasing trace lengths. (logarithmic scale)}
        \label{fig:discrete:runtime}
    \end{subfigure}
    \caption{Precision and runtime results for the discrete robot.}
    \label{fig:discrete}
\end{figure}
\emph{Discrete Robot.} 
As a first benchmark, consider the robot restricted to eight discrete directions from~\cite{DBLP:conf/rv/FinkbeinerFKK24}.
This example allows reproducing the results of the previous work using a fundamentally different implementation technique. Whereas~\cite{DBLP:conf/rv/FinkbeinerFKK24} translated \rlola specifications into standard Lola specifications, the results in~\Cref{fig:discrete} are obtained using the native symbolic implementation introduced in this paper.
In addition, it includes a comparison with UnaaLola~\cite{DBLP:journals/infsof/HiplerKLMSW26}.

The results are summarized in~\Cref{fig:discrete}, where~\Cref{fig:discrete:value} visualizes the absolute uncertainty range around the ground truth value of the robots x-position.
It becomes apparent that UnaaLola computes an interval approximation of the slack variables, while \rlola is able to compute a tighter bound.
\Cref{fig:discrete:runtime} reports runtimes for increasing trace lengths (note the logarithmic scale).
The SMT-based evaluation of UnaaLola is substantially slower than the presented implementation.
Compared to~\cite{DBLP:conf/rv/FinkbeinerFKK24}, the native implementation exhibits similar runtime tradeoffs between interval and symbolic evaluation while reducing the runtime overhead introduced by the earlier translation-based approach.

\begin{figure}[H]
    \centering
    \begin{subfigure}{0.49\linewidth}
        \includegraphics[width=\textwidth]{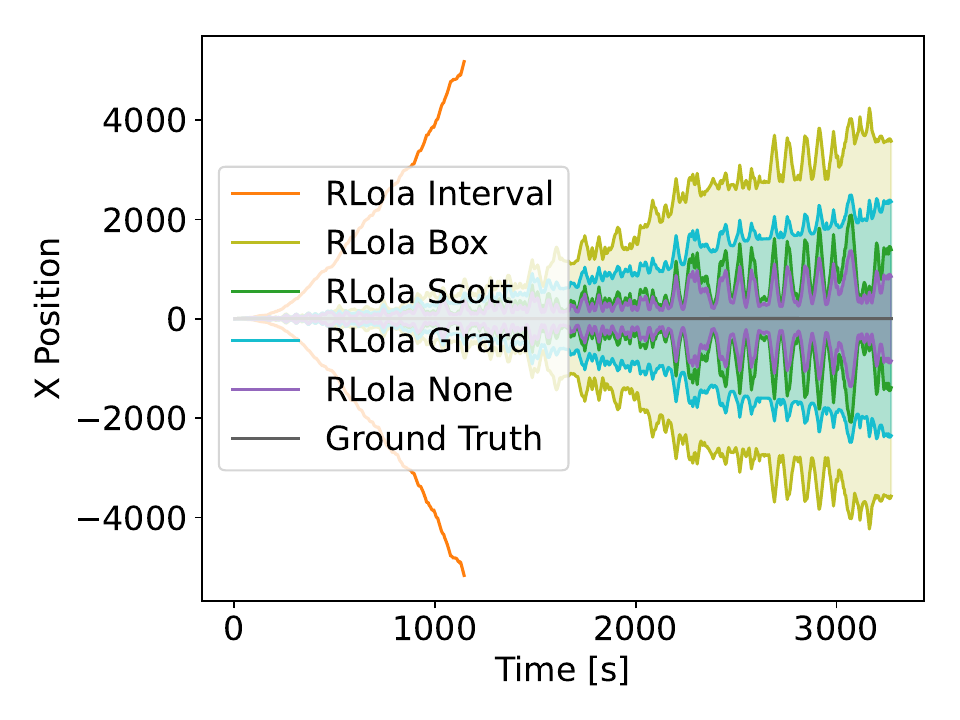}
        \caption{Range for the x-position over time.}
        \label{fig:omni_robot:time}
    \end{subfigure}
    \hfill
    \begin{subfigure}{0.49\linewidth}
        \includegraphics[width=\textwidth]{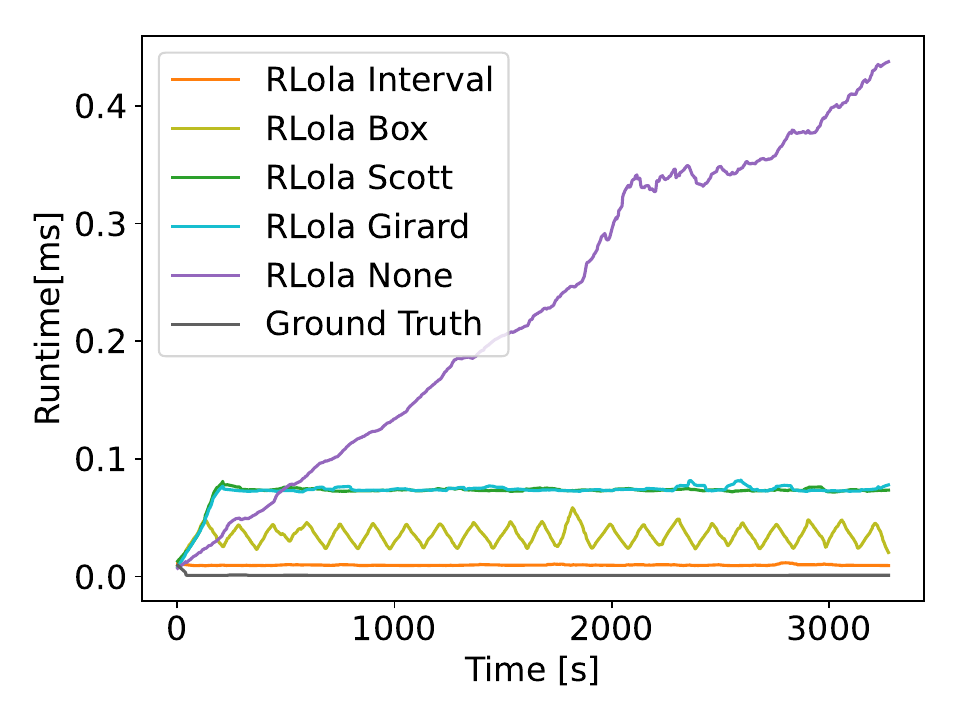}
        \caption{Runtime per method and time step.}
        \label{fig:omni_robot:limit}
    \end{subfigure}
        \begin{subfigure}{\linewidth}
        \centering
        \includegraphics[width=0.58\textwidth]{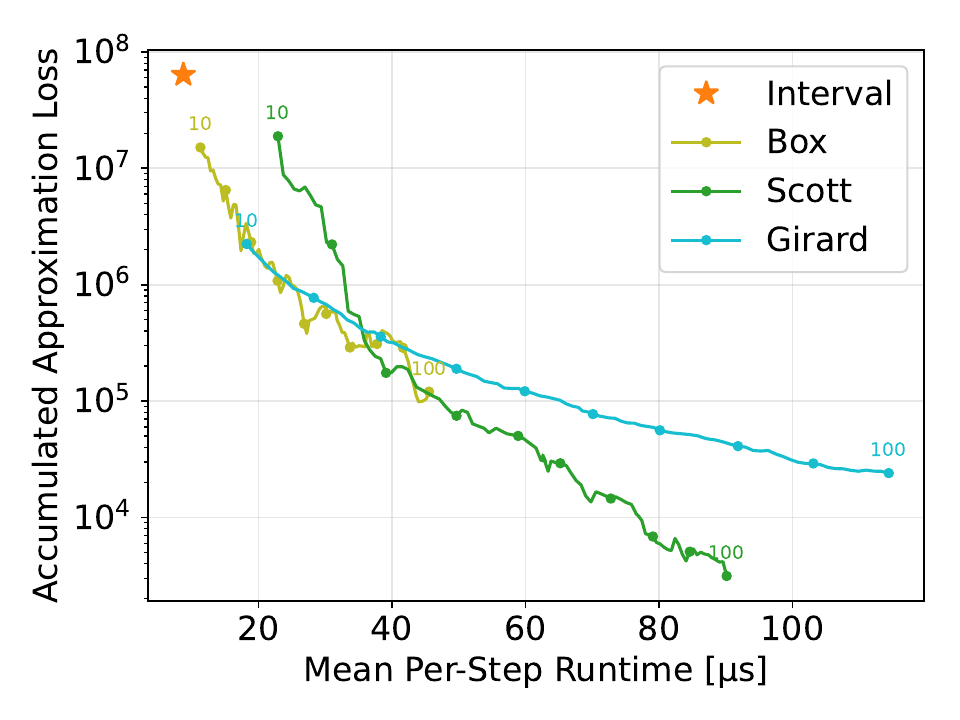}
        \caption{Trade-off between mean per-step runtime and accumulated approximation loss. Each curve is a single reduction method across bounds.}
        \label{fig:omni_robot:pareto}
    \end{subfigure}
    \caption{Precision and runtime for the omnidirectional moving robot.}
    \label{fig:omni_robot}
\end{figure}
\emph{Omnidirectional Robot.}
Compared to the previous benchmark, the motivating example presented in~\Cref{sec:motivation} allows movement in arbitrary direction.
Measurement noise contributes to both position streams with linearly independent coefficients, resulting in zonotopes that are difficult to approximate accurately.

\Cref{fig:omni_robot} summarizes the results for the monitor specified in~\Cref{spec:free_robot}.
\Cref{fig:omni_robot:limit} shows the absolute range of the $x$ position.
The box method performs an interval approximation whenever the given bound is exceeded, while the interval method is applied after each event no matter the bound.
\emph{\rlola None} never approximates and computes the optimal bounds using unbounded memory.
In this experiment, the slack variable bound is set to 50.
The Scott method~\cite{DBLP:journals/automatica/ScottRMB16} closely reproduces the precise (\rlola None) bounds.
The Girard method also tracks the precise bounds closely but introduces more approximation error than the Scott method.
As expected, interval arithmetic performs the worst.

\Cref{fig:omni_robot:time} shows the runtime behavior.
Without approximation, runtime grows linearly with the number of slack variables.
The Scott and Girard methods introduce additional overhead compared to simpler methods such as interval approximation, but all methods remain below 100 microseconds per event.

\Cref{fig:omni_robot:pareto} depicts the trade-off between precision, memory requirements, and runtime.
Each line in the graph represents a single reduction method evaluated at different generator bounds ranging from $10$ to $100$. The interval reduction is represented by a single point because it is independent of the generator bound. The x-axis measures the average per-step runtime, while the logarithmically scaled y-axis captures the accumulated approximation loss over complete traces, with both quantities averaged over 20 traces of length 200. Thus, a reduction method performs better the closer it is to the origin.

In general, the per-step runtime for each method increases with the generator bound, which is simply due to the additional computations required to maintain these generators.
Note that a higher generator count also increases the monitor's memory requirement.
The plot also shows that the loss decreases for higher generator counts, confirming the intuition that keeping more information about the true monitoring state leads to a higher precision.
Overall, the \emph{Girard} method shows diminishing returns rather quickly, while the precision of the \emph{Scott} method increases almost linearly with an increased generator count.

We were unable to obtain meaningful results for this benchmark using UnaaLola, likely due to numerical instabilities causing diverging bounds.
Partial results are included in~\Cref{fig:omni_robot_unaa} in the appendix.

\emph{Real Driving Emissions.}
Finally, we evaluate a real-world case study from~\cite{DBLP:conf/tacas/BiewerFHKSS21}, monitoring carbon monoxide emissions using onboard diagnostic data under realistic driving conditions.
We reproduce the experiments from~\cite{DBLP:journals/infsof/HiplerKLMSW26} using \rlola, varying the proportion of uncertain events from 1\% to 20\% in a trace of 5500 events and the noise amplitude from 1\% to 20\%.
\newpage
To model proportional and sporadic uncertainty, we exploit the quasi-constant nature of input stream variables at runtime, yielding the following encoding:
\begin{lstlisting}
input v_m: Float
input noise: Float
output e: Variable
output v := v_m + (noise/100.0) * v_m * e
\end{lstlisting}
The \stream{noise} input is set either to $0$ or to the selected noise amplitude, depending on whether the corresponding event is considered uncertain.
Whenever uncertain values occur in conditional expressions, we explicitly approximate them by taking the interval hull of the consequence and alternative:
\begin{lstlisting}
output s: Variable
output union := consequence.as_interval().union(alternative.as_interval())
...
if pAbove(condition, threshold) >= 1.0 then <consequence>
else if pBelow(condition, threshold) >= 1.0 then <alternative>
else union.center() + (union.width() / 2.0) * s
\end{lstlisting}
The final specification consists of 142 lines of code and is primarily concerned with determining whether the legal requirements for a \emph{realistic} drive are satisfied.
This is captured by the \stream{trip_valid} property.
For this benchmark, a trip is considered valid only if validity can be established with certainty.

\begin{figure}[H]
    \centering
    \begin{subfigure}{0.49\linewidth}
        \includegraphics[width=\textwidth]{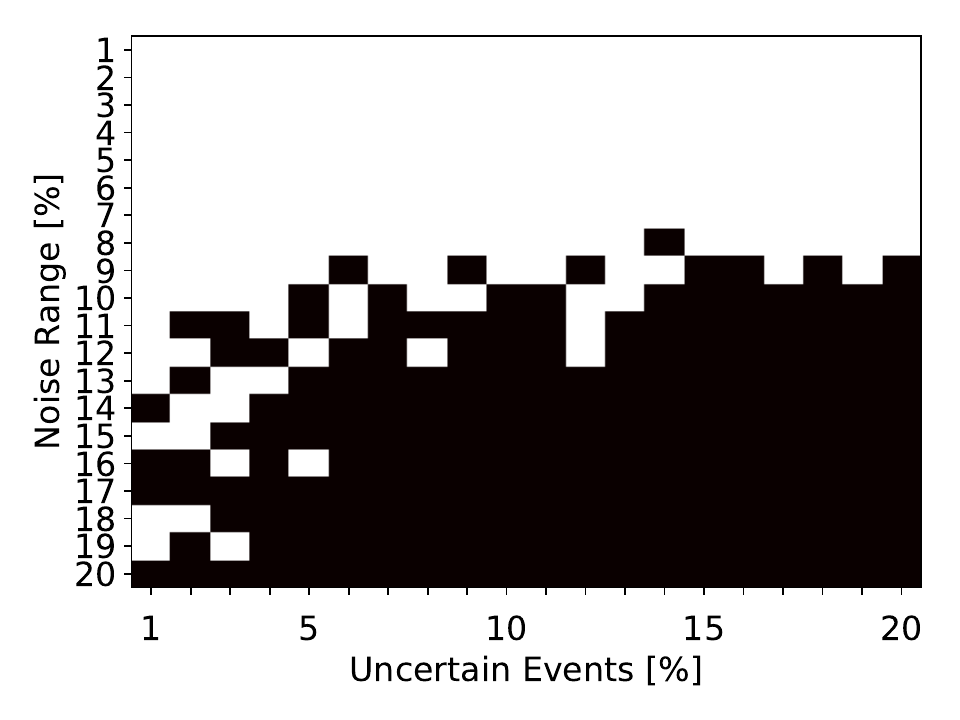}
        \caption{Trip-valid property monitored across different noise environments. Black cells indicate that the expected value was not computed.}
        \label{fig:rde:trip_valid}
    \end{subfigure}
    \hfill
    \begin{subfigure}{0.49\linewidth}
        \includegraphics[width=\textwidth]{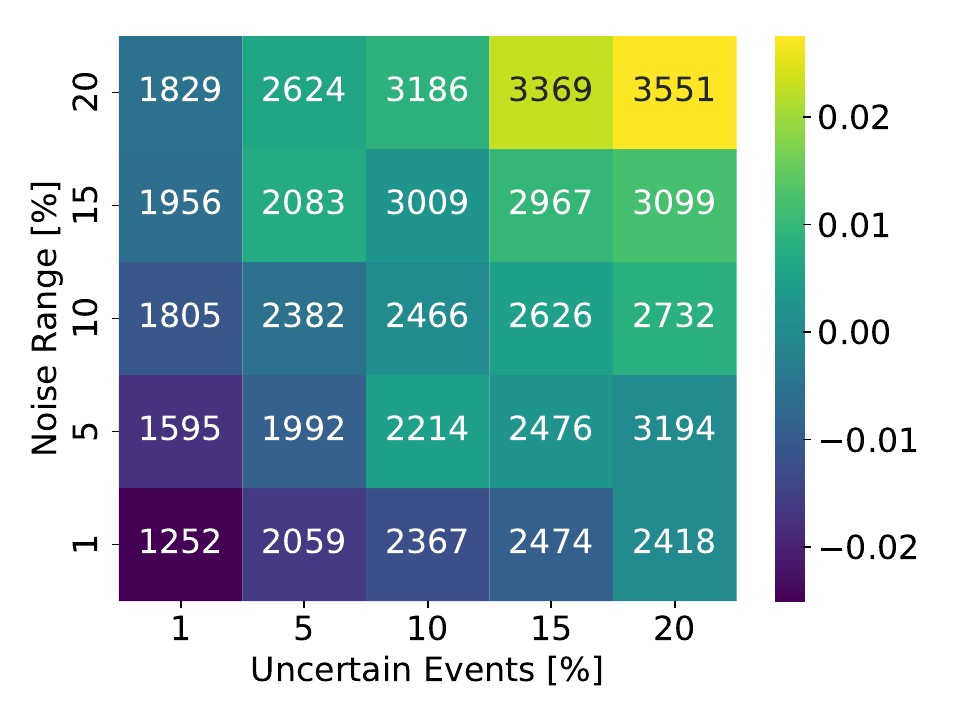}
        \caption{Relative runtime across different noise environments. Colors indicate {\rlola}’s runtime deviation from its average. Labels show the speedup compared to UnAALola.}
        \label{fig:rde:runtime}
    \end{subfigure}
    \caption{Trip-valid property and speedup for the RDE example.}
    \label{fig:rde}
\end{figure}

Black cells in~\Cref{fig:rde:trip_valid} indicate where \rlola cannot determine the expected verdict for the different noise environments.
The results match those obtained with UnAALola, although \rlola computes exact bounds while UnAALola relies on an approximate evaluator.
This suggests that the dominant source of uncertainty is the approximation of uncertain conditionals within the specification.
Further inspection reveals that the realistic trace is already close to violating the property, leaving limited tolerance for additional uncertainty.

\Cref{fig:rde:runtime} reports runtime results.
The heatmap shows \rlola{}’s runtime variance across environments, with an average runtime of $0.11$ seconds and a maximum variation of $0.02$ seconds.
For both \rlola and UnAALola, as noise increases, more symbolic variables are introduced, leading to increased runtime.
Cell labels show the speedup over UnAALola, whose runtime varies widely, likely due to the growth of Z3 constraints.

The evaluation answers \textbf{RQ2} and \textbf{RQ3} positively.
Compared to the state of the art, \rlola{}'s native symbolic evaluation achieves higher precision and substantially better performance.
At the same time, the case studies demonstrate that slack variables provide an expressive mechanism to model uncertainty in realistic systems without requiring domain-specific implementations.

%% file: conclusion.tex
\section{Conclusion}
We presented an online monitoring algorithm and interpreter for general \rlola specifications based on zonotopic abstractions.
The algorithm maintains a symbolic affine representation of the monitor state and tracks calibration and measurement errors through stream computations. Zonotope over-approximations enable sound bounded-memory monitoring as slack variables accumulate.

We evaluated several approximation strategies with respect to approximation error and false-positive rates.
The results show that the choice of approximation method significantly affects both precision and runtime, with the Scott method providing the best overall trade-off on our benchmarks.

We implemented the approach in the \rlola monitoring framework as a runtime monitoring tool for arbitrary \rlola specifications.
The implementation combines precise pruning for the bounded-memory fragment with zonotopic approximations for the general case.
Our evaluation on synthetic and real-world benchmarks demonstrates substantial improvements in precision and runtime compared to SMT-based approaches.

Future work includes developing approximation strategies tailored specifically to runtime monitoring.
Unlike existing zonotope reduction methods from reachability analysis, such strategies should account for long-term error propagation through future stream computations and trigger evaluations.

%% file: appendix.tex
\newpage
\appendix
\section{Appendix}
\subsection{Additional Evaluation}
\begin{figure}[H]
    \centering
    \begin{subfigure}{0.49\linewidth}
        \includegraphics[width=\textwidth]{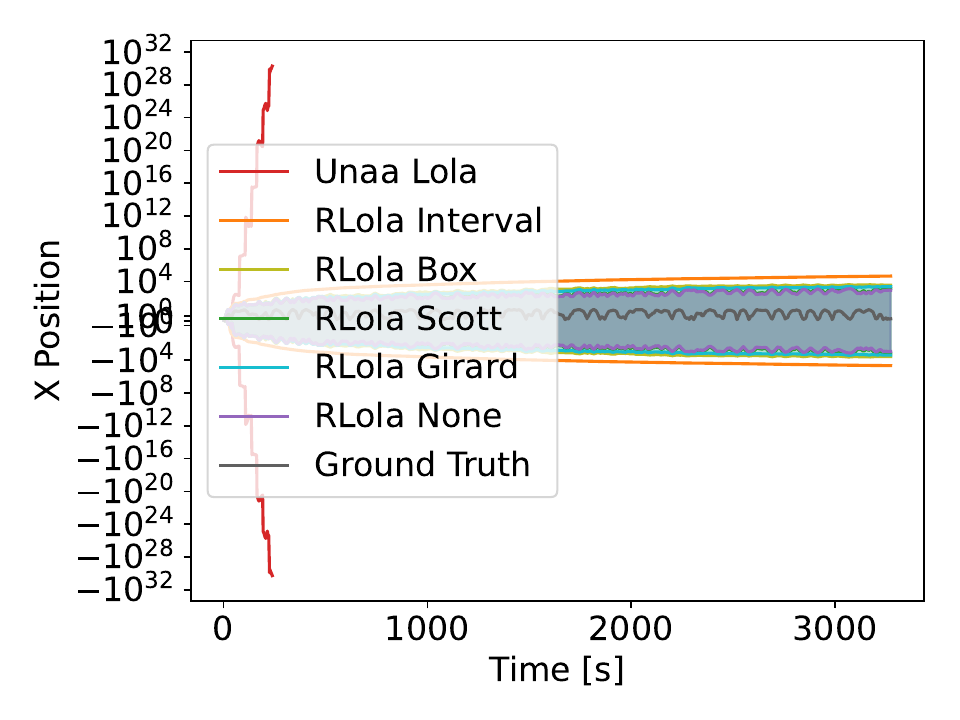}
        \caption{Estimated range for the x-position over time.}
        \label{fig:omni_robot_unaa:time}
    \end{subfigure}
    \hfill
    \begin{subfigure}{0.49\linewidth}
        \includegraphics[width=\textwidth]{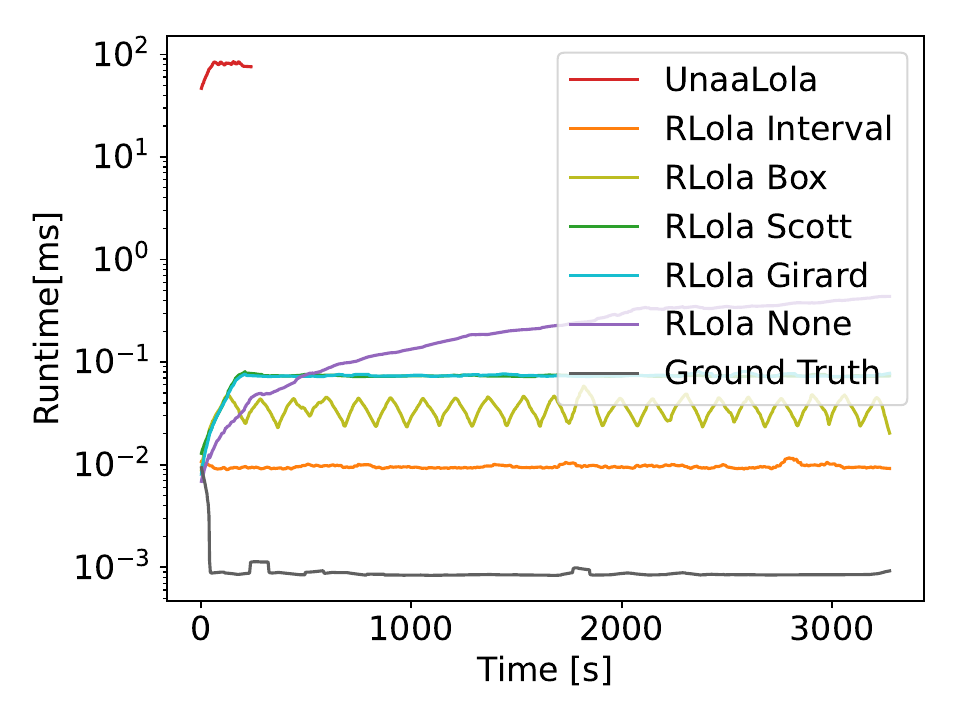}
        \caption{\rlola tool runtime per approximation method per time step.}
        \label{fig:omni_robot_unaa:limit}
    \end{subfigure}
    \caption{Precision and runtime results for the freely moving robot.}
    \label{fig:omni_robot_unaa}
\end{figure}